\documentclass[namedreferences]{SolarPhysics}
\usepackage{spr-sola-addons} 
\usepackage{url}             

\usepackage[pdftex]{graphicx} 

\newcommand{\degr}{{\hbox{$^\circ$}}}

\newcommand{\arcsec}{{\hbox{$^{\prime\prime}$}}}
\renewcommand{\sun}{_\odot}

\newcommand{\be}{\begin{equation}}
\newcommand{\ee}{\end{equation}}
\newcommand{\bea}{\begin{eqnarray}}
\newcommand{\eea}{\end{eqnarray}}
\newcommand{\beas}{\begin{eqnarray*}}
\newcommand{\eeas}{\end{eqnarray*}}

\renewcommand{\tabcolsep}{0.14cm}

\newcommand{\aap}{    {\it Astron. Astrophys.}}

\newcommand{\apj}{    {\it Astrophys. J.}}

\chardef\us=`\_ 

\begin{document}

\begin{article}

\begin{opening}

\title{Structure of the transition region and the low corona from TRACE and SDO observations near the limb}

\author[addressref={aff1},corref,email={calissan@cc.uoi.gr}]{\inits{C.E.}\fnm{C. E.}~\lnm{Alissandrakis}}
\author[addressref={aff1}]{\inits{A.}\fnm{A.}~\lnm{Valentino}}

\address[id=aff1]{Section of Astro-Geophysics, Department of Physics, University of Ioannina, GR-45110 Ioannina, Greece}

\runningtitle{Structure of the Transition Region and Low Corona }

\begin{abstract}
We examined the structure near the solar limb in TRACE images of the continuum and in the 1600 and 171 \AA\ bands as well as in SDO images in the continuum (from HMI) and all AIA bands. The images in different wavelength bands were carefully coaligned  by using the position of Mercury for TRACE and Venus for SDO during their transit in front of the solar disk in 1999 and 2012 respectively. Chromospheric absorbing structures in the TRACE 171\,\AA\ band are best visible 7\arcsec\ above the white light limb, very close to the inner limb, defined as the inflection point of the rising part of the center-to-limb intensity variation. They are correlated with, but are not identical to spicules in emission, seen in the 1600\,\AA\ band. Similar results were obtained from AIA and SOT images. Tall spicules in 304\,\AA\ are not associated with any absorption in the higher temperature bands. Performing azimuthal averaging of the intensity over 15\degr\ sectors near the N, S, E and W limbs, we measured the height of the limb and of the peak intensity in all AIA bands. We found that the inner limb height in the transition region AIA bands increases with wavelength, consistent with a bound-free origin of the absorption from neutral H and He. From that we computed the column density and the density of neutral hydrogen as a function of height. We estimated a height of $(2300\pm500)$\,km for the base of the transition region. Finally, we measured the scale height of the AIA emission of the corona and associated it with the temperature; we deduced a value of  $(1.24\pm0.25)\times10^6$\,K for the polar corona

\end{abstract}
\end{opening}

\section{Introduction} \label{intro} 
Homogeneous one-dimensional models of the solar atmosphere put the base of the Chromosphere-Corona Transition Region (TR) at a height of about 2\,000\,km ({\it i.e.} a mere 3\arcsec) above the $\tau_{\rm5000}=1$ level ({\it e.g.} \citealp{1981ApJS...45..635V, 1990ApJ...355..700F, 1991ApJ...377..712F, 1993ApJ...406..319F, 2008ApJS..175..229A, 2015ApJ...811...87A}). This is well below the height of spicules, which may exceed 15\arcsec\ ({\it e.g.} \citealp{2018SoPh..293...20A}).

The observational evidence for the position of the start of the TR came originally from eclipse data and is related to relative timing between the instance that the lunar disk covers the photosphere and the instance at which coronal lines appear after the occultation of the chromosphere by the moon \citep{1966soat.book.....Z}. More recently, \cite{2008ApJS..175..229A} placed the top of the chromosphere at 2140\,km, roughly consistent with the maximum intensity of the 10833\,\AA\ He{\sc i} and the the 5875\,\AA\ D$_3$ lines. On the other hand, microwave observations place the start of the TR above sunspots as low as 1500\,km ({\it e.g.} \citealp{2019SoPh..294...23A}). Microwave observations near the limb give important information on the TR structure. In this range the opacity, and hence the limb height, increases with wavelength, but the situation is complicated by the low resolution of full-disk observations and the presence of spicules.

Direct measurements beyond the limb in the extreme UV and the X-rays are difficult, due to the low spatial resolution of the older space-born instruments. \cite{1995ApJ...453..929D} used the {\it Normal Incidence X-ray Telescope} (NIXT) to measure the position of the limb at 63.5\,\AA\ with a spatial resolution of $\sim1$\arcsec, while \cite{1998ApJ...504L.127Z} used the {\it Extreme-Ultraviolet Imaging Telescope} (EIT) aboard the {\it Solar and Heliospheric Observatory}, (SoHO) at 171, 195, 284 and 304\,\AA\ with $\sim3$\arcsec\ resolution On the other hand, although both the {\it Transition Region and Coronal Explorer} (TRACE) and the {\it Solar Dynamics Observatory} (SDO) have sufficient resolution (of the order of than 1\arcsec), the measurement requires very precise relative pointing among white light observations and observations in the EUV channels.

Beyond the photospheric limb, spicules are well visible in the TRACE UV images \citep{2005ESASP.600E..54A} and in the SDO 1600\,\AA\ band images, although the normal exposure time is rather short in the latter case; they are also seen as absorbing features in the EUV images of both instruments, giving a rugged appearance to the lower boundary of the corona near the limb. However, no systematic study of spicules in absorption has been published so far.
 
In this work we used the position of Mercury and Venus, transiting the solar disk and observed by TRACE and SDO respectively, to make accurate alignment of UV/EUV images with white light images. Subsequently we made a comparative study of limb features in emission (UV bands) and in absorption (EUV bands); for the Venus transit we also compared such features with structures in the Ca{\sc ii} H-line using images from the {\it Solar Optical Telescope} (SOT) on board Hinode. Furthermore, we computed the average radial variation of the intensity as a function of distance from the photospheric limb, and from that the height of the limb and of the peak intensity for each wavelength band. We also measured the intensity scale height of the coronal emission, which is a very important tool for the computation of the coronal temperature. In Section \ref{section:obs} we describe the data and our methodology, in Section \ref{section:res} we present the results of our analysis. We summarize our conclusions in Section \ref{section:disc}.

\section{Observations, Data Processing and Pointing Corrections}\label{section:obs}

\begin{figure}
\begin{center}
\includegraphics[width=0.8\textwidth]{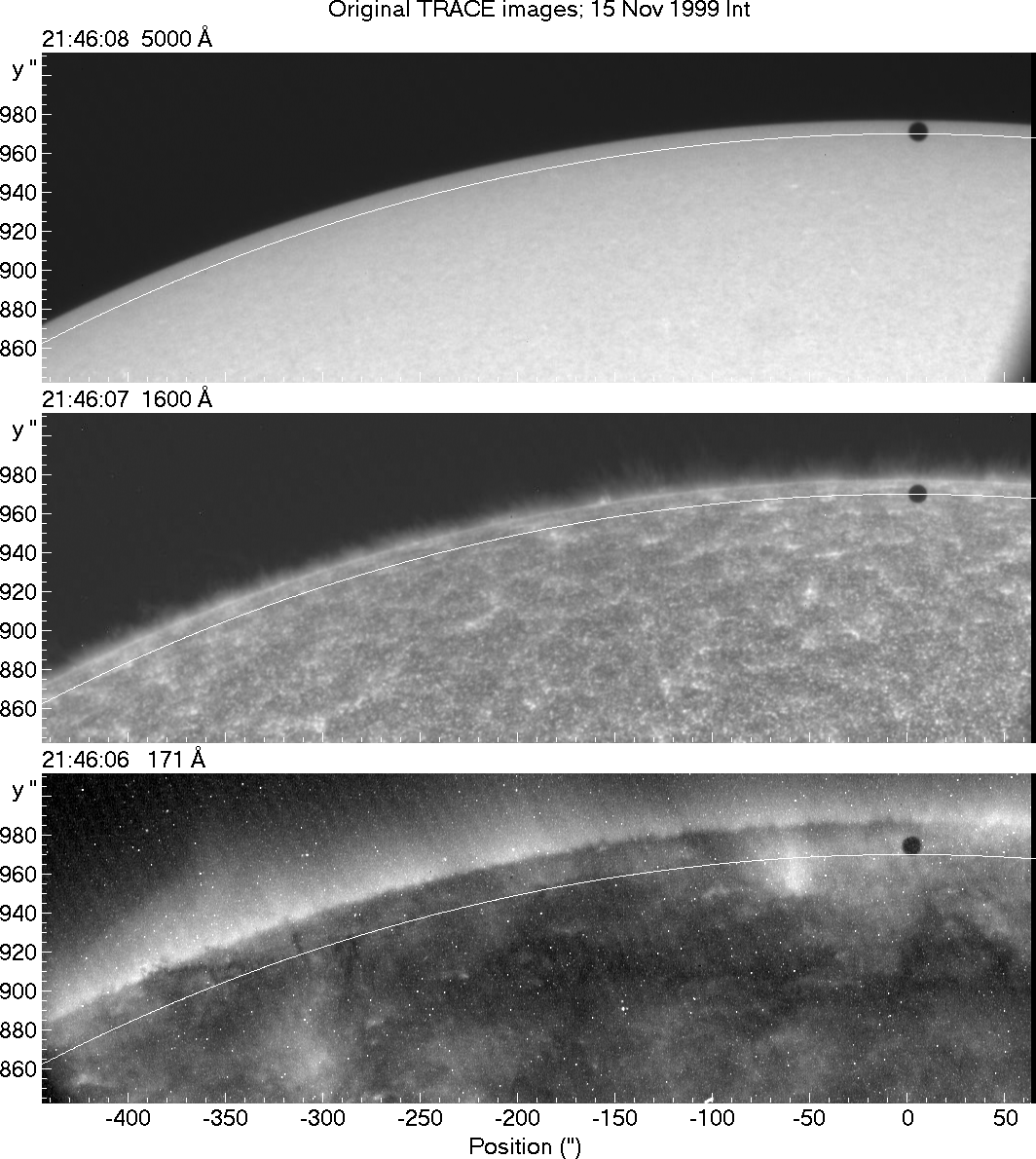}
\end{center}
\caption{Original images of Mercury transiting the solar disk in white light, the 1600\,\AA\ band and the 171\,\AA\ band. The white arc shows the position of the limb according to the pointing in the image file header.}
\label{fig:org}
\end{figure}

\subsection{The 1999 Transit of Mercury Observed by TRACE}\label{section:Merc}
TRACE observed the transit of Mercury on November 15 1999 from 21:08 to 21:56 UT in three wavelength bands: white light (WL), 1600\,\AA\ and 171\,\AA. The trajectory of the planet on the solar disk, as viewed from TRACE, was near the north pole (Figure \ref{fig:org}), almost in the EW direction, the northern limb of Mercury being very close to the photospheric limb at central meridian passage. The Yohkoh {\it Soft X-ray Telescope} (SXT) images show a quiet region devoid of coronal holes at that location. The first contact in WL was at about 21:19 UT, at a position angle of 8.3\degr, while Mercury was visible in front of the corona at 171 \AA\ from the beginning of the image sequence. The observations were interrupted near 21:56 UT, when TRACE entered the shadow of the Earth; at that time Mercury was located 65\arcsec\ W of the central meridian. Images were taken sequentially in the three bands with a cadence of 38\,s; exposure times were 0.0032, 4.10 and 19.48\,s for WL, 1600 and 171 bands respectively. In addition to the usual 1024 by 1024 pixel field of view, TRACE provided uncompressed 1024 by 384 pixel images; we used those for the 171\,\AA\ band, to avoid artifacts due to the compression.

As can be seen in Figure \ref{fig:org}, the 171\,\AA\ band images suffered from hot pixels; these were removed by comparing each image with the previous and the next image. In addition, these images showed diagonal, regularly spaced, striations which were partly removed by appropriate spatial filtering. We note that spicules are well visible in the 1600 images, while spicule-like features appear in absorption at the base of the corona in the 171 band images.

The white arc in Figure \ref{fig:org} shows the position of photospheric limb, according to the information provided in the headers of the image FITS files. It is more than clear that the pointing is considerably off in all three bands. In order to correct the pointing we started from the WL images. We computed the gradient of each image and determined the position of the limb from its maximum value. Subsequently we performed a least square fit of the positions to a circle, determining in this way the center of the solar disk and its radius; the WL limb been very regular, the fit was very good, with an root mean square (rms) deviation of $\sim0.05$ pixel. 

We noticed that the radius determined by the fit was not constant during the time sequence, but varied by $\pm0.15$ \%, with a periodicity of $\sim20$ min. We could not identify the origin of this variation, in particular we did not detect any correlation with the temperature of TRACE components recorded in the FITS headers; moreover, a fit of the limb in the 1600 band did not show a similar variation of radius. We thus decided to do the fit using a constant solar radius, equal to the average for the sequence, which was 973.4\arcsec, assuming the nominal pixel size of 0.5\arcsec; this corresponds to 715\,km per \arcsec. We note that the ephemeris gives a value of the radius equal to 970.26\arcsec; the difference could be attributed to a deviation of the pixel size from the nominal value by 0.03\%, but it is most probably due to the fact that the arc of the solar limb in our images is only $\sim30$\degr\ long and, consequently, its radius is not determined with high accuracy, although the position of the limb is much more accurately measured. In any case, these differences are too small to affect the accuracy of our results.

\begin{figure}[h]
\begin{center}
\includegraphics[width=0.8\textwidth]{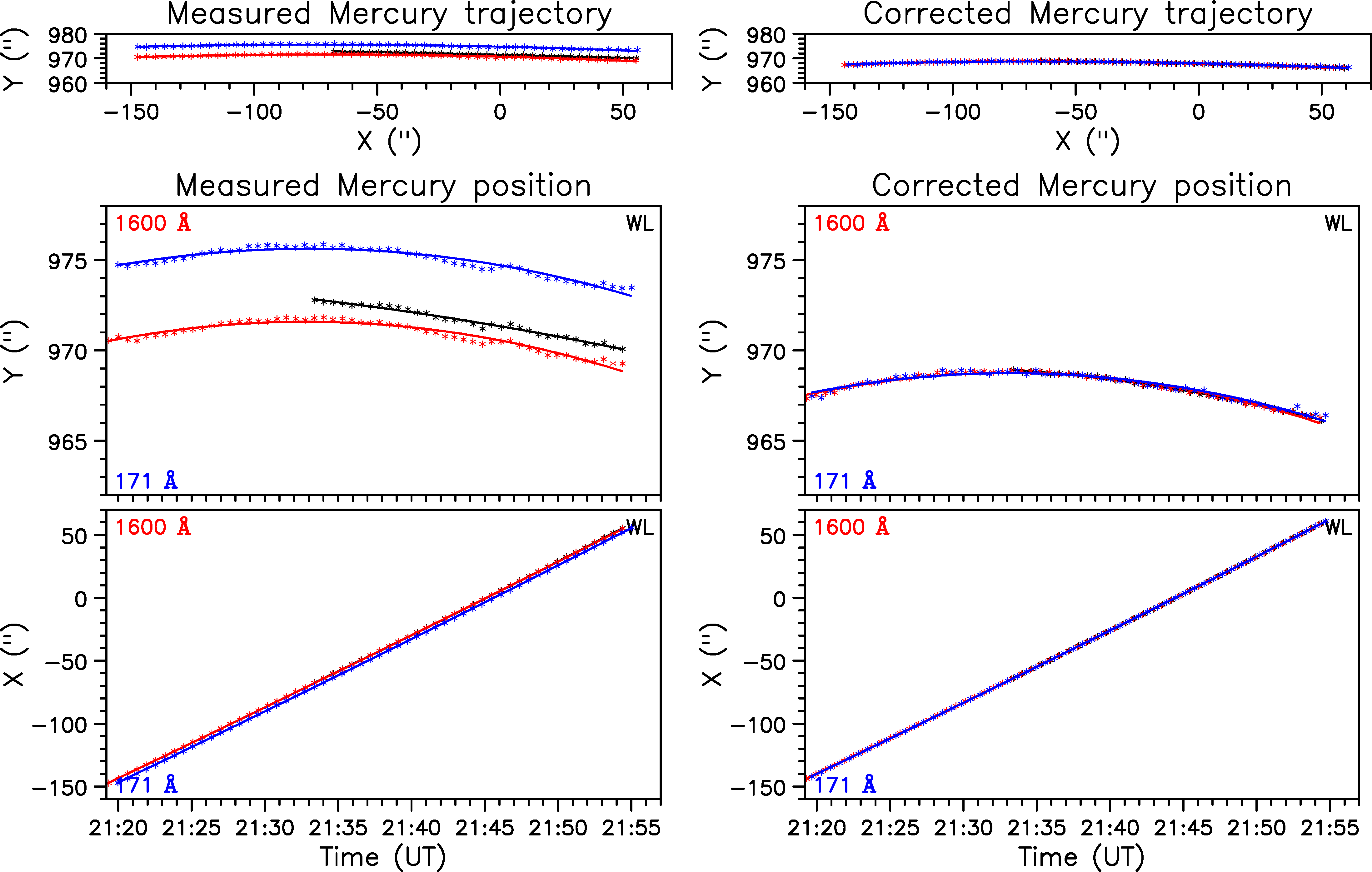}
\end{center}
\caption{Trajectory and position of Mercury on the solar disk as  a function of time before (left) and after (right) pointing correction. Full lines show a second degree fit.}
\label{fig:pos}
\end{figure}

Having established the position of the limb for WL images, we measured the position of Mercury. This was done again by computing the gradient of the images to get the border of the planet's disk and determining by least square fit the position of its center and its radius for all three wavelength bands. Obviously, this was possible only when a sufficient part of the disk of Mercury was visible, hence WL measurements could not be done before 21:34 UT. At this point we should like to note that we found practically the same value for the radius of Mercury for all images and all bands, $4.93\pm0.04$\arcsec, which shows that the image scale did not vary among the three wavelength bands; the ephemeris value of 4.97\arcsec\ is within the error margin of our measurements.

\begin{table}[h]
\caption{Pointing corrections, $\Delta x, \Delta y$, for TRACE in pixels}
\label{Table01}
\begin{tabular}{lcc}
\hline 
Band  &\cite{1999SoPh..187..229H} & This work       \\
\hline
1600  &$~~\,1.2,-1.2$             &  $~~\,0.5, -1.9$\\
171   &$-4.3, ~~\,7.2$            &  $-2.5, ~~\,6.3$\\
\hline 
\end{tabular}
\end{table}    

The pointing corrections were determined from the difference of the position of Mercury between the 1600 and 171 bands on the one hand and the WL band on the other, after interpolation of the positions in 1600 and 171 to the time of the nearest WL image, under the assumption that images in all thee bands have the same orientation. The corrected trajectory and position of Mercury \begin{figure}[h]
\begin{center}
\includegraphics[width=.9\textwidth]{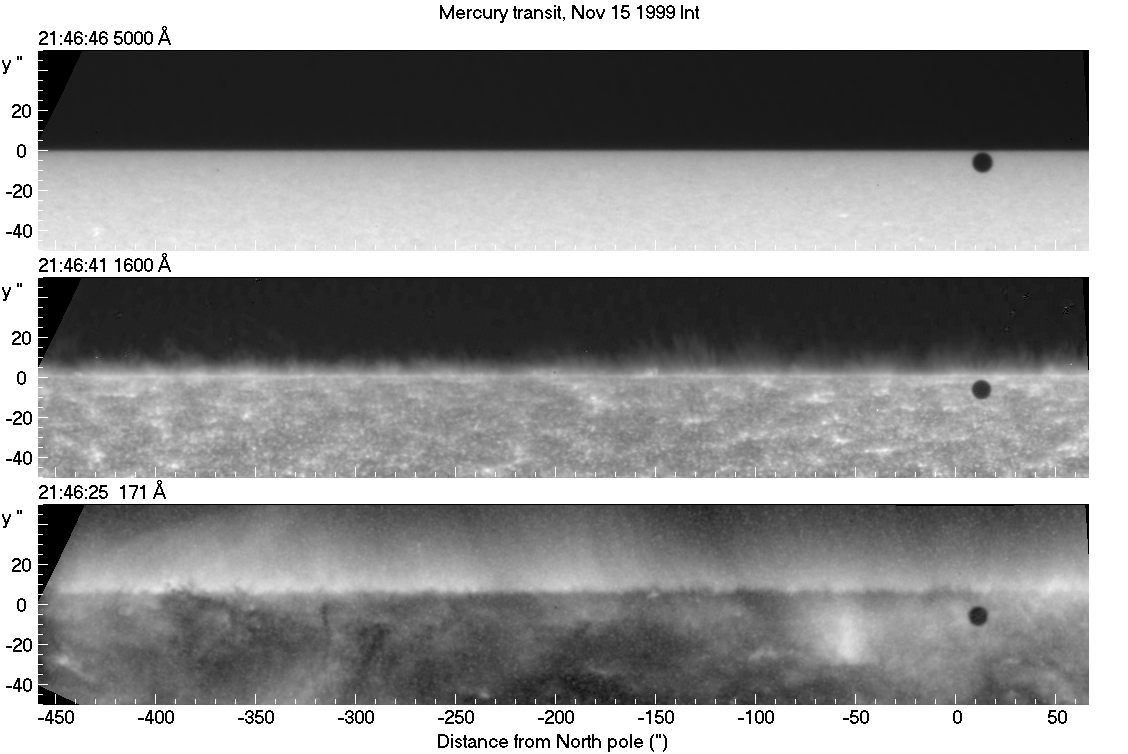}
\end{center}
\caption{A frame from Movie 1, showing polar projection images of Mercury transiting the solar disk, after pointing correction. The 171 \AA\ image has also been corrected for hot pixels and partly corrected for diagonal striations.}
\label{fig:cor}
\end{figure}
are shown in the plots in the right panel of Figure \ref{fig:pos}, together with the original measurements (left panel), from which it is obvious that the pointing correction is excellent. Our corrections with respect to the WL pointing are given in Table \ref{Table01}, together with those from \cite{1999SoPh..187..229H}. We note that the difference between the two sets is from 0.7 to 1.8 pixels; we add for completeness that the offset of the true WL pointing from that on the header of the FITS file was $\Delta x=7.5, ~ \Delta y= 7.5$ pixels. 

The images shown in Figure \ref{fig:org} are shown again in Figure \ref{fig:cor}, this time in polar projection so that the limb shows as a straight horizontal line, while the entire sequence is included in Movie 1. In addition to the pointing correction, the 171 \AA\ image has been corrected for hot pixels and striations. 

\subsection{The 2012 Transit of Venus Observed by SDO and SOT}\label{section:Venus}
\begin{figure}[h]
\begin{center}
\includegraphics[width=\textwidth]{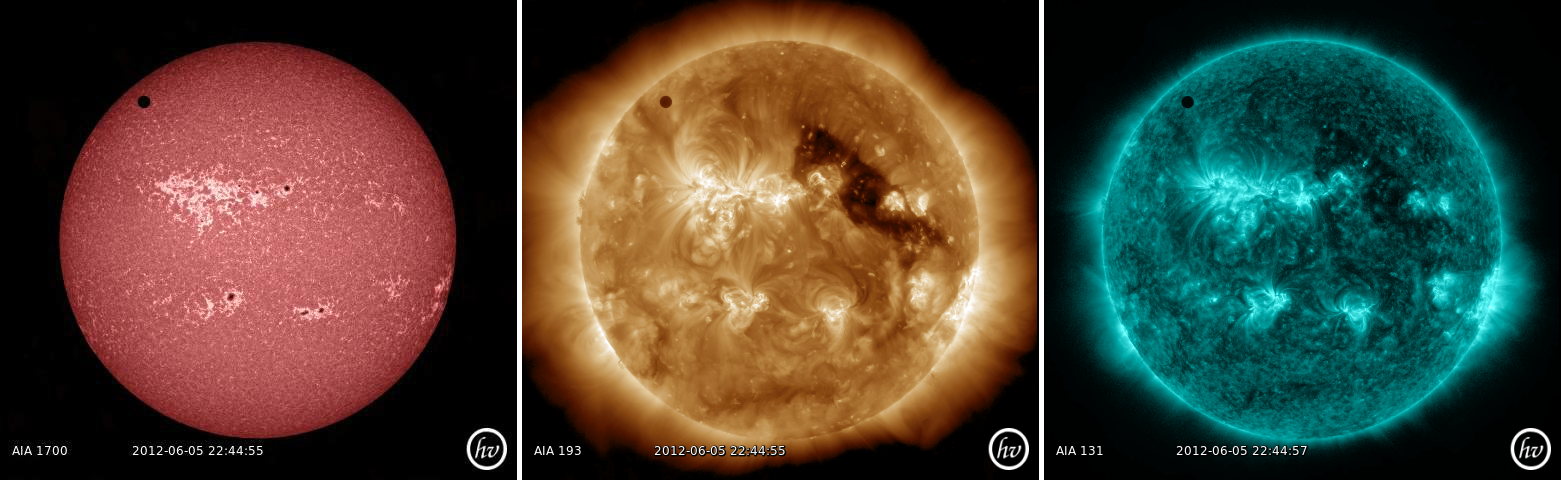}
\end{center}
\caption{Full disk AIA images in the 1700, 171 and 131\,\AA\ bands from https://helioviewer.org/ near the beginning of the Venus transit, at 22:45 UT of 5 Jun 2012. The planet shows up as a dark disk in the NW quadrant.}
\label{fig:FD_AIA}
\end{figure}

For SDO the fist contact of Venus with the solar disk was near 22:28 UT of June 5, 2012 at a position angle of 46\degr\ and the fourth around 04:35 UT of the next day at a position angle of $-60$\degr. We selected two five-minute intervals for further processing, one shortly after the second contact, around 22:45 UT (Figure \ref{fig:FD_AIA}) and one before the third contact, around 04:00 UT. The image cadence was 12\,s for the six EUV bands of the {\it Atmospheric Imaging Assembly} (AIA) (94, 131, 193, 211, 304, and 335\,\AA) and 24\,s for the two UV bands (1600 and 1700\,AA). The exposure time was 1.0\,s for the 1700\,\AA\ band, 2.0\,s for the 171 and 193\,\AA\ bands and 2.9\,s for all others. The cadence of the {\it Helioseismic and Magnetic Imager} (HMI) images was 45\,s. The image scale was $\sim0.6$\arcsec\ per pixel for AIA and 0.5\arcsec\ per pixel for HMI.

We measured the position of the center of the solar disk in the HMI images and the position of the Venus center in the HMI and the AIA images in the way described in Section \ref{section:Merc}. Due to the short exposure time, the images in some AIA bands were quite noisy. In such cases we smoothed the images by convolution with a 2\arcsec\ Gaussian; this gave a reliable measurement of the Venusian disk, without affecting its position or radius. The pointing corrections were computed by comparing the position of Venus in the AIA images with that on the HMI images; having two widely separated time intervals we were also able to check the scale and orientation of the AIA images, assuming that the corresponding HMI quantities were exact.

The fit to a circle of the solar limb, determined from the gradient of the HMI images, was excellent with an rms deviation of 0.03\arcsec (0.06 pixels), close to the value of $0.050\pm0.026$ given by \cite{2012SoPh..275..261W} for the HMI image distortion. The deviation is mostly due to slight departure of the SDO image from a perfect circular disk, with an amplitude of 0.1\arcsec\ (0.2 pixels); this gives an estimate of the accuracy of the position of the WL limb. 

We found that the offset of the HMI images center was less than 0.3 pixels (0.15\arcsec) from the values recorded in the image file header, and the solar radius 1.00023 that of the header value. Our method is accurate enough to reveal the slight decrease, by 0.25\arcsec, of solar radius between the start and the end of the transit. A similar fit of the 1700\,\AA\ AIA image gradient gave a larger RMS, 0.2\arcsec, due to the less regular shape of the limb at that wavelength and a radius  1.21\arcsec\ greater than that of the WL; for the 1600\,\AA\ band we obtained a limb position 1.66\arcsec\ above the WL limb, while for the other bands a fit of the limb was impossible to to the irregularities.

The rms of the fit of the Venusian disk to a circle ranged from 0.1\arcsec\ for the 171, 193 and 211\,\AA\ bands to 0.3\arcsec\ for the 94\,\AA\ band, reflecting the noise level of the respective images. We found an average radius of 29.4\arcsec, slightly higher than the ephemeris value of 28.9\arcsec, obtained from https://ssd.jpl.nasa.gov/horizons.cgi for SDO. We also found a systematic decrease of the radius with wavelength, from 29.6\arcsec\ at 94\,\AA\  to 29.3\arcsec\ at 6193\,\AA\ (HMI); given the accuracy of our measurements, this effect is probably real and may reflect higher absorption of the Venusian atmosphere at short wavelengths (see also \citealp{2015NatCo...6E7563R}).

\begin{table}[h]
\caption{SDO Pointing corrections}
\label{Table:SDOcorr}
\begin{tabular}{rrrrrr}
\hline 
Tel &\multicolumn{1}{c}{$\lambda$}&\multicolumn{1}{c}{$F_s$}&\multicolumn{1}{c}{$\Delta\theta$}&\multicolumn{1}{c}{$\Delta X_0$}&\multicolumn{1}{c}{$\Delta Y_0$}\\
    &\multicolumn{1}{c}{\AA}      &                     &\multicolumn{1}{c}{\degr}         & pixels                         & pixels                         \\
\hline 
  - & 6173 & 1.00000 &    0.0000 & $-$0.01 & $-$0.49 \\
 \hline 
 1 &  131 & 0.99995 & $-$0.0109 & $-$0.57 &    0.55 \\
  1 &  335 & 1.00002 & $-$0.0023 & $-$0.60 & $-$0.47 \\
\hline 
  2 &  193 & 0.99983 &    0.0012 & $-$0.09 &    0.26 \\
  2 &  211 & 0.99991 &    0.0007 & $-$0.32 & $-$0.71 \\
\hline 
  3 & 1700 & 0.99988 & $-$0.0065 &    0.00 &    0.00 \\
  3 & 1600 & 0.99984 & $-$0.0010 & $-$0.10 &    0.10 \\
  3 &  171 & 0.99984 & $-$0.0017 & $-$0.12 &    0.59 \\
\hline 
  4 &  304 & 1.00010 & $-$0.0038 & $-$0.34 & $-$0.63 \\
  4 &   94 & 1.00011 &    0.0079 & $-$0.24 &    0.37 \\
\hline 
\end{tabular}
\end{table}    

Our computed pointing corrections are given in Table \ref{Table:SDOcorr}, where the first column gives the AIA telescope number, $F_s$ is the scale correction factor, $\Delta\theta$ the correction of the image orientation and $\Delta X_0, \Delta Y_0$ the corrections of the position of the center of the solar disk, all with respect to the corresponding values in the image file headers. The most important source of error is the accuracy of measurement of the position of Venus, thus from the rms deviation of the fit given above we expect these corrections to be good to about 0.1\arcsec. 

The corrections are minute and can be ignored for all practical purposes. For example, the range of scale corrections corresponds to a radial shift of 0.26\arcsec\ at the limb, and the range of orientation corrections to a similar azimuthal shift. Still, in order to have the highest possible accuracy, we preferred to apply these corrections in the subsequent analysis of our data. 

The Transit of Venus was also observed by SOT, on board Hinode. We used SOT images in the wavelength band of the Ca{\sc ii} H-line, close to our selected AIA images, to compare spicular structure. Their centering was performed by comparison with disk features seen near the limb in the UV AIA images. 

\section{Results}\label{section:res}

\subsection{Chromospheric Spicules and TR Absorption Features}\label{spic}
\begin{figure}[h]
\begin{center}
\includegraphics[width=.85\textwidth]{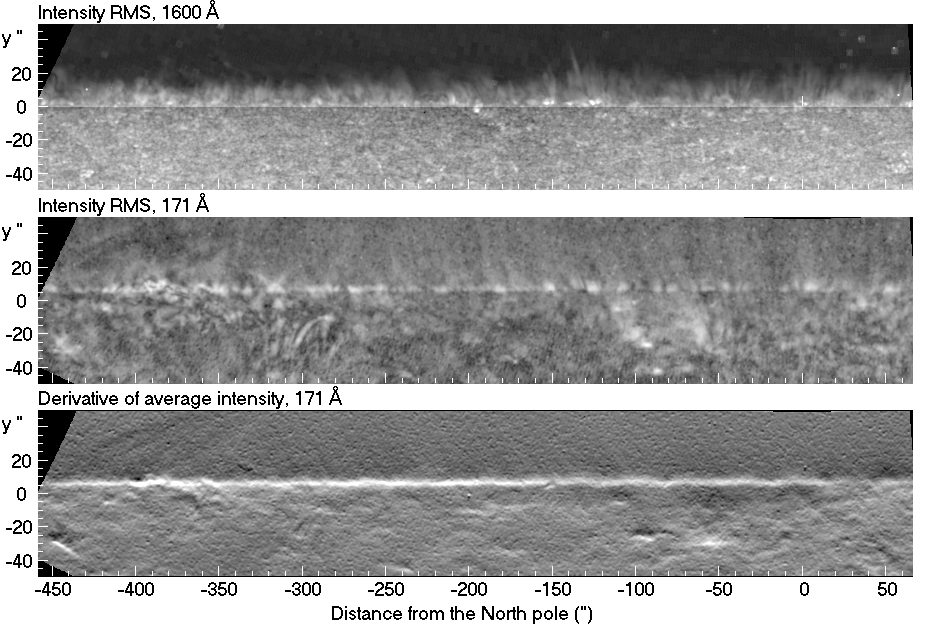}
\end{center}
\caption{Top: Image of the intensity rms in the 1600\,\AA\ band. Middle: Same for the 171\,\AA\ band. Bottom: The derivative of the intensity in the radial direction for the 171\,\AA\ band. For the 1600 rms image we have used different display scales above and below the limb. The images are shown in polar projection, as in Figure \ref{fig:cor}.}
\label{fig:spicules_rms}
\end{figure}

We will start our discussion with the TRACE 1600\,\AA\ band. As shown in Figure \ref{fig:cor}, we have limb brightening which peaks right above the WL limb. Beyond that spicules are well visible, as noted above. Their time variability is shown well in Movie 1, as well as in the image of the intensity rms, computed over the 35\,min interval of our observations and shown in the top panel of Figure \ref{fig:spicules_rms}. In both figures most spicules extend up to 15\arcsec\ (10\,800\,km) beyond the limb, while individual features go up to $\sim30$\arcsec\ (21\,000\,km). We note that no spicular structures are visible below the limb, either in individual images or in the intensity rms.

In the 171\,\AA\ band the chromosphere is completely opaque at low heights, as evidenced from the fact that no individual absorption features are visible there in the image of Figure \ref{fig:cor}. Such features are seen in the height range from $\sim5.5$\arcsec\ up to $\sim10$\arcsec, best visible in the rms image of Figure \ref{fig:spicules_rms} (middle panel). Their variability peaks around 7\arcsec, very near the height at which the radial derivative of the 171 intensity maximizes (bottom panel of Figure \ref{fig:spicules_rms}).

A close look at the bottom panel of Figure \ref{fig:spicules_rms} reveals that the white band marking the position of the 171 limb is slightly inclined, by 0.15\degr. This effect could be due to a difference in orientation between the 171 and the other bands, for which we could not check. If it were real, it would imply that the 171\,\AA\ limb is higher near the North pole, the regression analysis giving a height of 4.9\,Mm at the pole and 4.0\,Mm at the position angle  of 8.3\degr.

\begin{figure}[h]
\begin{center}
\includegraphics[width=.49\textwidth]{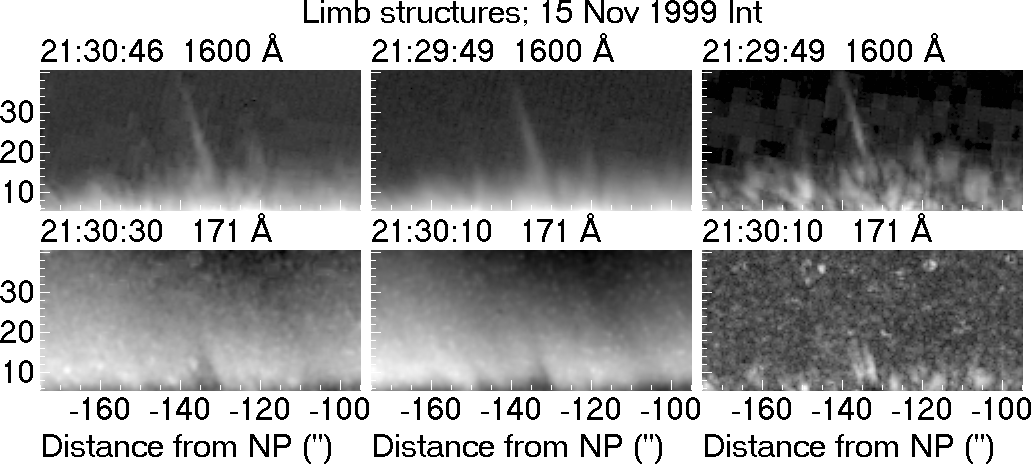}~\includegraphics[width=.49\textwidth]{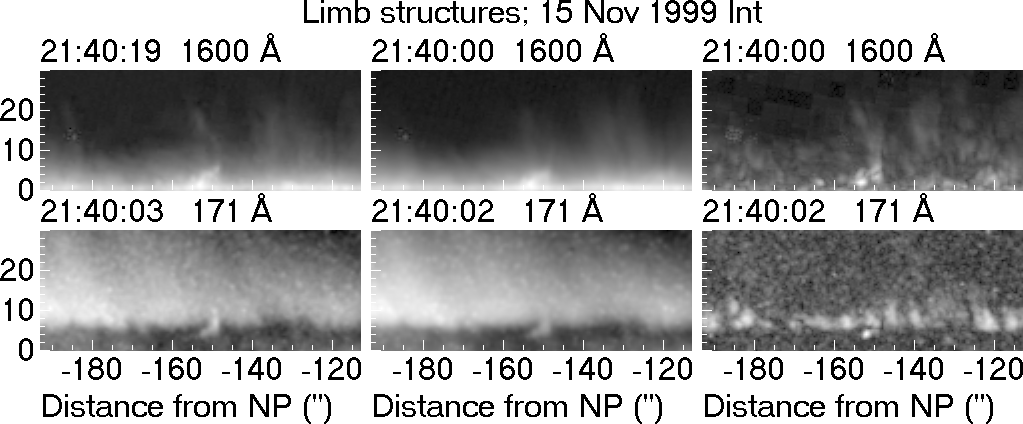}
\end{center}
\caption{Left set: A jet-like feature in the 1600\,\AA\ (top) and the 171\,\AA\ bands; in addition to the image near the peak of the feature (left) we give images of the average intensity (middle) and the rms (right) over 10\,min. Right set: Same for a bright feature.}
\label{fig:feature2}
\end{figure}

In Figure \ref{fig:feature2} we identify two features in the 1600 and 171\,\AA\ bands. The first is a jet-like feature which lasted about 10 min and extended beyond 30\arcsec, shown in the left part of the figure. In order to reduce noise we give, in addition to an image near its maximum, images of the average intensity and its rms, computed over 10 min. The feature is much more prominent in 1600, but its lower part is still detectable in 171. The second feature (right in Figure \ref{fig:feature2}) is a transient brightening which, judging from the 1600\,\AA\ images (see Movie 1), had its base beyond the limb. It is visible in the 171\,\AA\ band from a hight of $~\sim4$\arcsec, indicating that its emission is completely absorbed below that.

\begin{figure}[h]
\begin{center}
\includegraphics[width=\textwidth]{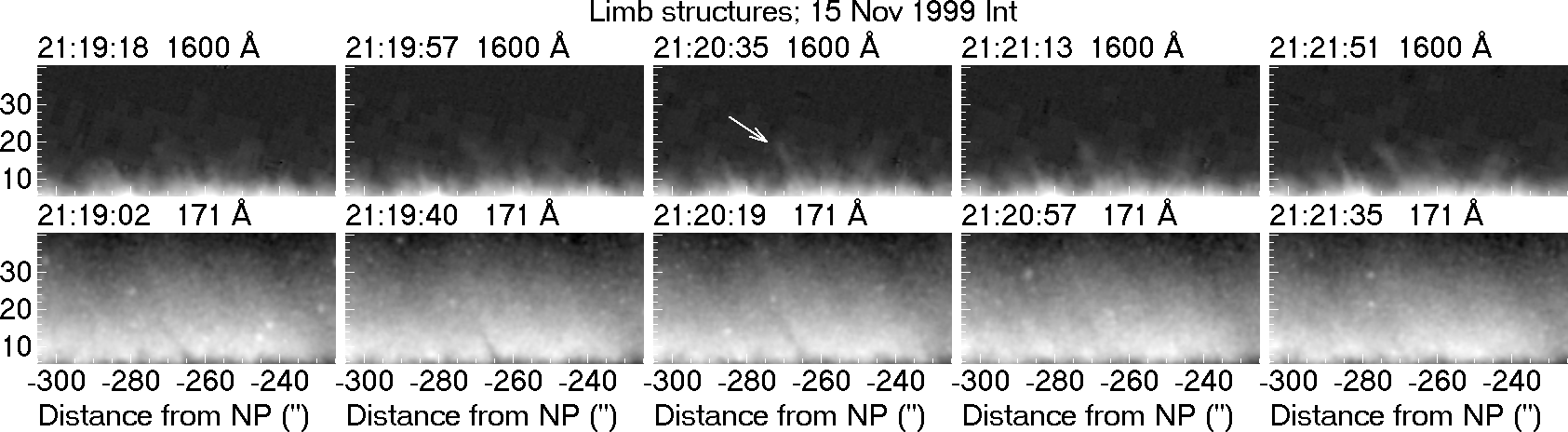}
\end{center}
\caption{A five image sequence of a jet-like feature (arrow) in the 1600\,\AA\ (top) and the 171\,\AA\  bands.}
\label{fig:feature1}
\end{figure}

A second jet-like feature is shown in Figure \ref{fig:feature1}. Here it is interesting to note that, although the coincidence between the two bands is fairly good (middle image), the feature appeared and disappeared first in the 171 and then in the 1600\,\AA\ band.

\begin{figure}[h]
\begin{center}
\includegraphics[width=\textwidth]{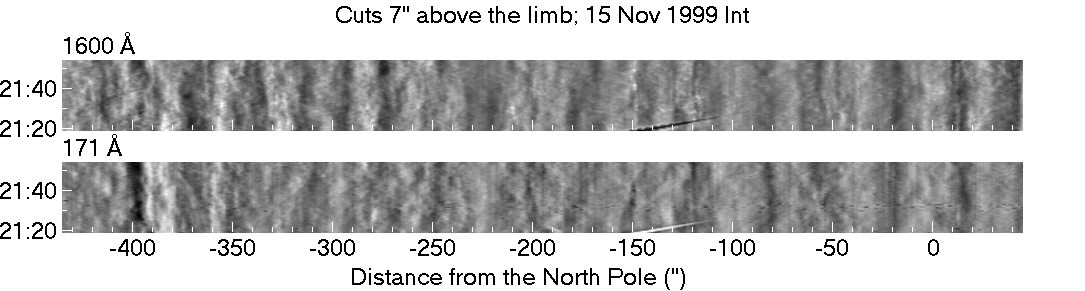}
\end{center}
\caption{Images (cuts) of the intensity as a function of position ans time at a height of 7\arcsec\ above the limb in the 1600\,\AA\ band (top) and in the 171\,\AA\ band (bottom, negative image). A 20\arcsec\ high-pass filter has been applied to enhance the visibility of small-scale features. The inclined dark (white) line in the lower part of the images near $-150$\arcsec\ is is due to the image of Mercury.}
\label{fig:cut1}
\end{figure}

In order to make a more complete comparison, we computed the intensity as a function of position and time for a 2\arcsec\ wide zone at fixed heights for both bands. Two such cuts, for 1600\,\AA\ and 171\,\AA, are shown in Figure \ref{fig:cut1}, at a height of 7\arcsec\ where the variability of the 171 emission is strongest; the 171\,\AA\ cut is displayed as a negative to facilitate the comparison with the 1600\,\AA\ image. The two cuts are similar, indicating the the absorption features are spicule-associated. The overall correlation coefficient is rather low, $-0.41$, and the differences between the images do not allow a reliable computation of the time delay between the two wavelength bands.

The AIA images during the Venus transit gave us the opportunity to compare limb structures in many spectral bands and all over the limb. We note that during the transit the south limb was not recorded in all wavelength bands, therefore we used images before the transit, around 20:27 UT,  integrated over 1\,min  to improve the signal to noise ratio. We also deconvolved the EUV images by the instrumental responses (Point Spread Function, PSF) computed by \cite{2013ApJ...765..144P}, to reduce stray light beyond the limb.

\begin{figure}[h]
\begin{center}
\includegraphics[width=.95\textwidth]{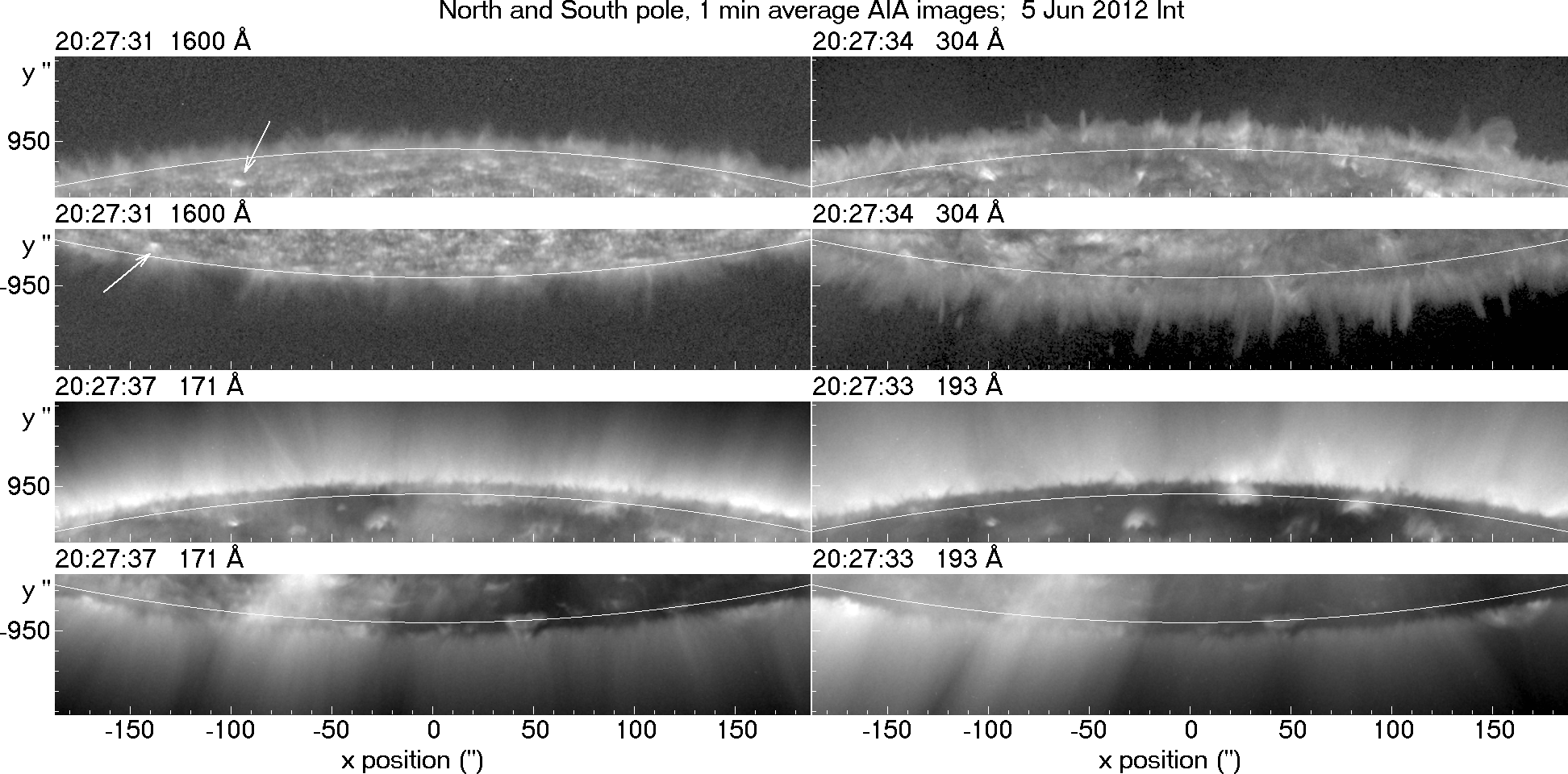}
\end{center}
\caption{Comparison of polar regions in the 1600, 304, 171 and 193\,\AA\ AIA bands. The white arc marks the photospheric limb. The 1600\,\AA\ images have been corrected for center-to-limb variation by subtracting 90\% of the azimuthally averaged intensity. The EUV images have been deconvolved with the PSF of \cite{2013ApJ...765..144P}. The arrows point to 1600\,\AA\ features near the limb, which are also visible at 304\,\AA.}
\label{fig:limb_AIA}
\end{figure}

The best regions were around the N and S poles, as there was some activity in the E and W limbs (Figure \ref{fig:FD_AIA}). Images around the north and south pole are presented in Figure \ref{fig:limb_AIA} in the 1600, 304, 171 and 193\,\AA\ AIA bands, which form at chromospheric, low TR, upper TR and coronal temperatures respectively.

\begin{figure}
\begin{center}
\includegraphics[width=\textwidth]{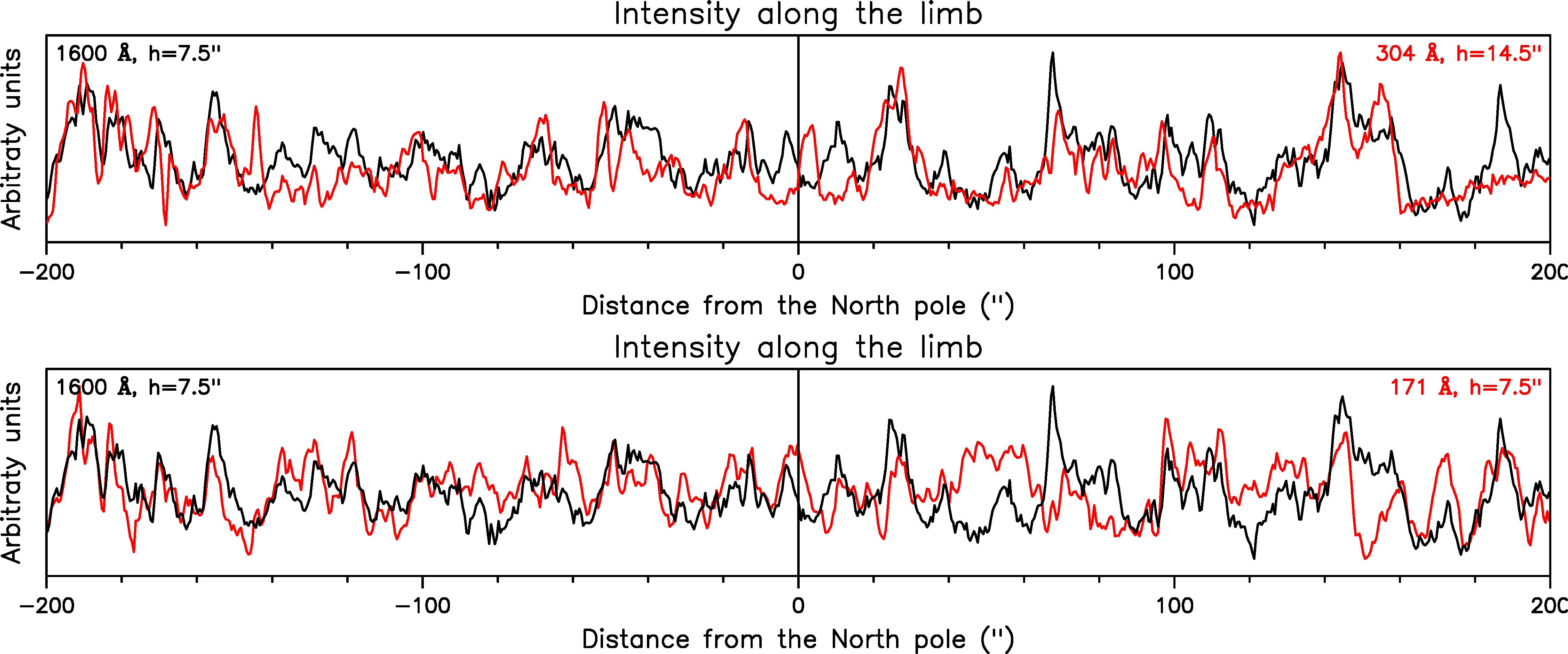}
\end{center}
\caption{Top: Intensity tracings parallel to the north limb in the 1600\,\AA\ band (7.5\arcsec\ above the limb, black curve) and the 304\,\AA\ bands (14.5\arcsec\ above the limb, red curve). Bottom: Similar tracings for the 1600 and 171\,\AA\ bands, both at a hight of 7.5\arcsec above the limb. For 171\,\AA\ the negative of the intensity is plotted. All tracings have been treated with a 50\arcsec\ high-pass spatial filter to reduce large-scale variations.}
\label{fig:IntLimb}
\end{figure}

As expected, spicules are more prominent and higher in the 304\,\AA\ band than in the 1600\,\AA\ band ({\it e.g.} \citealp{1999A&A...341..610G}). The plot in the upper panel of of Figure \ref{fig:IntLimb} shows that the 1600 intensity 7.5\arcsec\ above the north polar limb is very similar to the 304 intensity 7\arcsec\ higher.  In the 1600 and 304\,\AA\ bands spicules extend much higher in the S pole than in the north; although Figures \ref{fig:FD_AIA} and \ref{fig:limb_AIA} show no coronal hole there, the examination of AIA images of previous days showed that the south polar region was at or near the boundary of coronal hole which, on the day of the transit, was on the back side of the solar disk. This is probably the reason why 1600\,\AA\ spicules are weaker, though taller, in the south than in the north.We note further that disk structures, such as the ones marked by arrows in the figure, appear closer to the limb in the 1600 than in the 304 band, verifying that the latter forms higher in the atmosphere, as reported by \cite{2018A&A...619L...6N}.

Images in the higher temperature AIA bands show the usual absorption above the limb. The similarity of individual absorption features at 171\,\AA\ with emission features at 1600\,\AA\ is demonstrated in the lower plot of Figure \ref{fig:IntLimb}. Both curves are 7.5\arcsec\ above the WL limb. We note, however, that there is no trace of the big 304\,\AA\ spicules, particularly near the S limb, either in the 171 or the 193\,\AA\ images (Figure \ref{fig:limb_AIA}), which shows that these structures do not absorb the EUV radiation.

\begin{figure}[!h]
\begin{center}
\includegraphics[width=\textwidth]{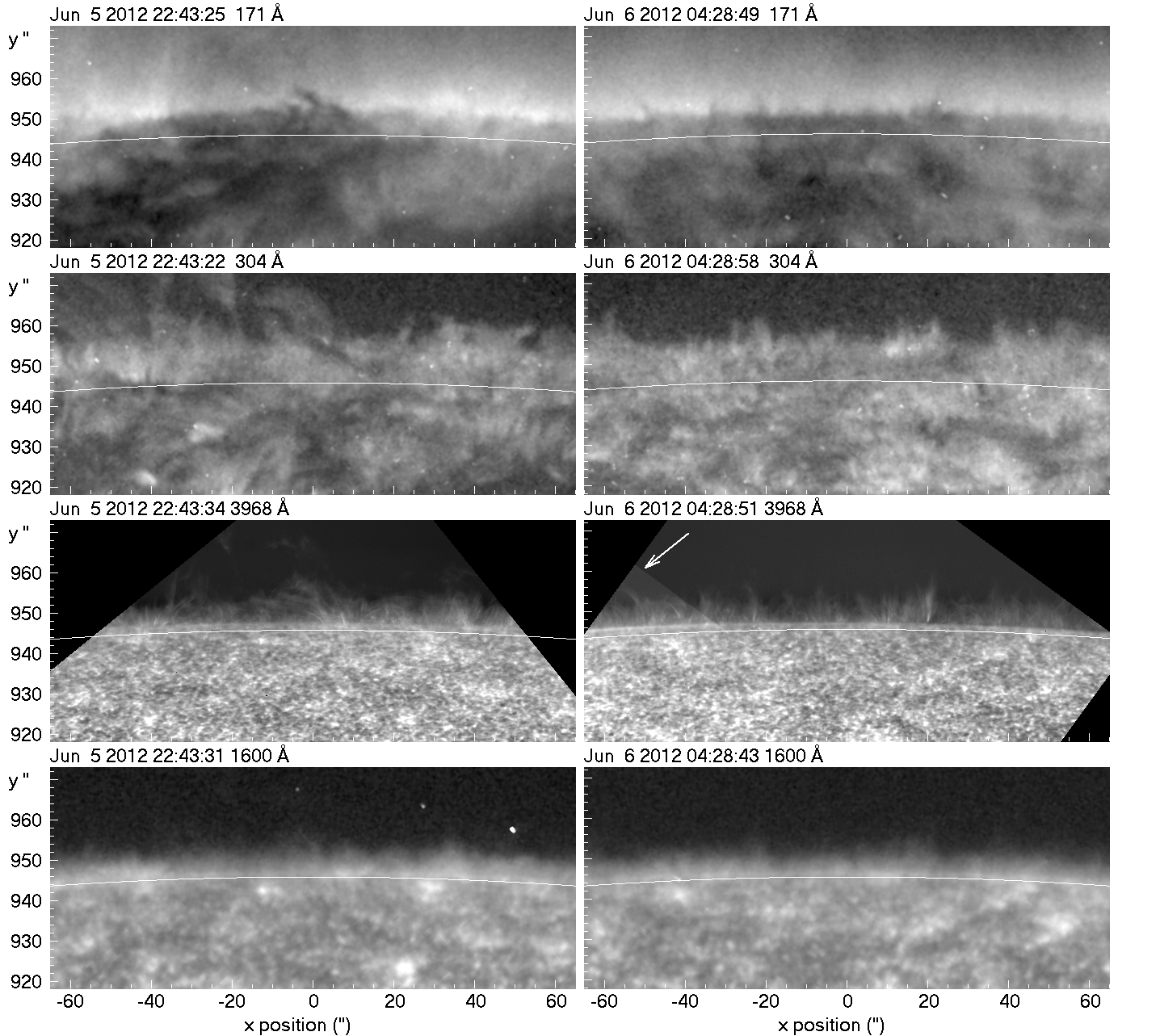}
\end{center}
\caption{AIA images in the 171, 304 and 1600\,\AA\ bands and SOT images in the 3968\,\AA\ (Ca{\sc ii} H-line) band, after the start (left) and before the end (right)  of the transit. The images are oriented so that the photospheric limb (white arc) is in the horizontal direction. H-line and 1600\,\AA\ images have been corrected for center-to-limb variation. The straight-line feature marked by the arrow is an artifact.}
\label{fig:limb_SOT}
\end{figure}

Figure \ref{fig:limb_SOT} compares SOT images in the Ca{\sc ii} H-line with AIA images in the 171, 304 and 1600\,\AA\ bands, close to the second (left) and third (right) contact, when both Venus and the limb were in the SOT field of view. There is some activity in the region shown in the left column of the figure, but the one shown at right is quiet. In spite of the lower resolution and the higher noise in the AIA images, most spicules in the H-line are identified with emission features in the 1600 and 304\,\AA\ bands and with absorption features in the 171\,\AA\ band.

\subsection{Radial Intensity Profiles Near the Limb}\label{radprof}
\begin{figure}[h]
\begin{center}
\includegraphics[width=\textwidth]{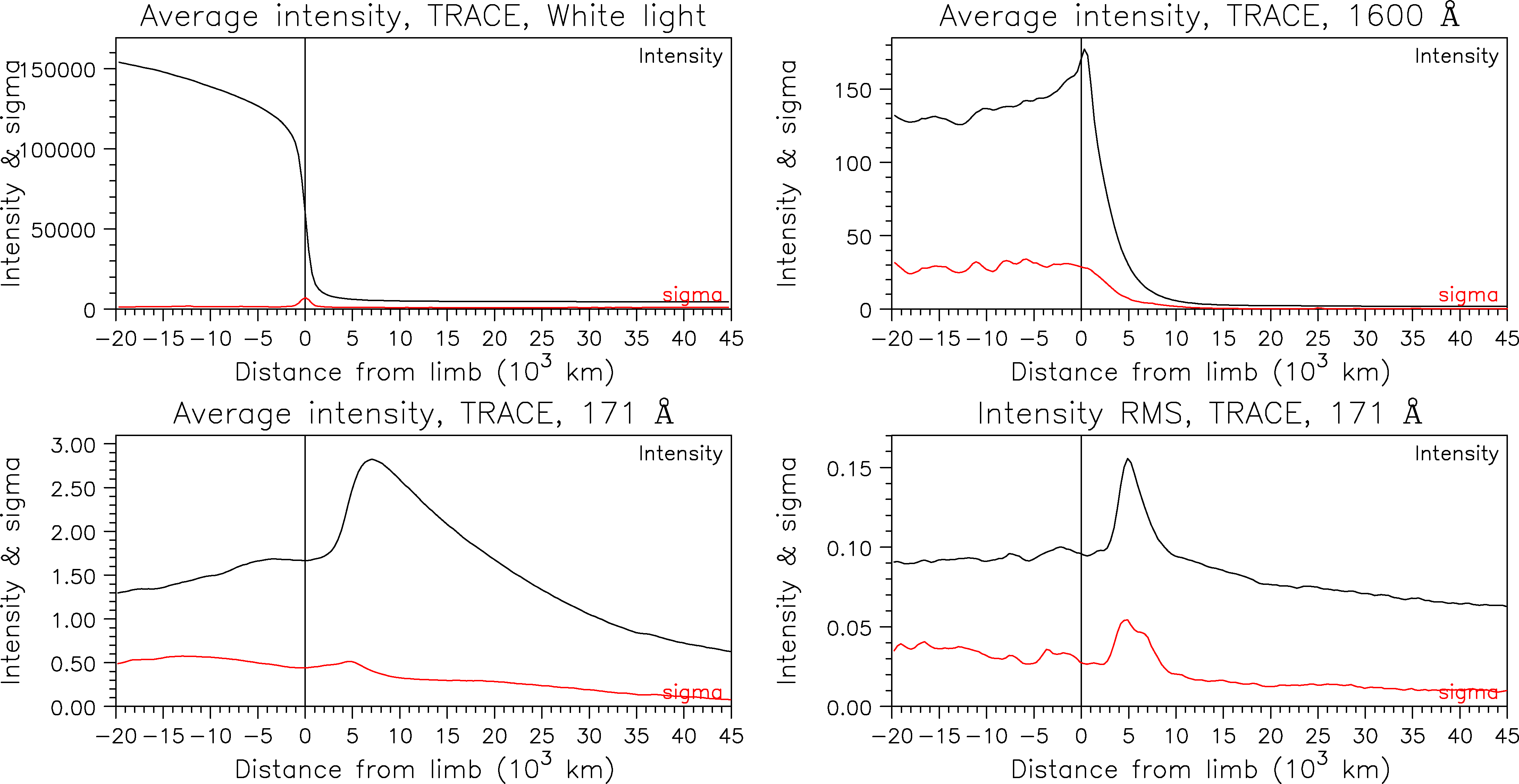}
\end{center}
\caption{The azimuthally averaged intensity, as a function of distance from the white-light limb for the three TRACE wave bands and the intensity rms of the 171\,\AA\ band. Curves in red show the rms of the respective average values, which is a measure of the azimuthal intensity variation at each radial position.}
\label{fig:cnl}
\end{figure}

We computed the radial intensity variation, by integrating over position angle. For TRACE we used the full field of view, avoiding regions affected by vignetting, and the average intensity for the WL, the 1600\,\AA\ and 171\,\AA\ bands are shown in Figure \ref{fig:cnl}, together with the respective rms. In the same figure we give a similar plot for the 171 intensity rms image (middle panel of Figure \ref{fig:spicules_rms}). The distance scale is in Mm from the WL limb.

We first note that there is no change in the 171\,\AA\ intensity or rms at the location of the WL limb; this proves that, just above the limb, we have indeed absorption by chromosphere material of the coronal emission from behind the plane of the sky, rather than lack of emission in a region void of coronal plasma. We note further that, as expected from the images in Figure \ref{fig:cor}, the 171 curve shows a sharp rise starting at a height of about 4.4\,Mm; the intensity peaks around 8.2\,Mm and an exponential decrease follows. We define the height of the 171 limb as the position of the inflection point in the rising part of the curve; this height is slightly smaller, by 0.5\arcsec, than the height at which the 171 rms peaks. These heights apparently represent the height at which the spicule forest becomes optically thin and radiation from behind the plane of the sky becomes visible. We also measured the height of the 1600 limb from the inflection point of the decaying part of corresponding curve. These values are given in Table \ref{Table02}, together with our measurements of the intensity scale height; we note that the latter may be affected by stray light, for which we have no information.

\begin{table}[h]
\caption{Limb parameters from TRACE observations}
\label{Table02}
\begin{tabular}{lccc}
\hline 
$~\,\lambda$  & Limb height &Peak height & Intensity Scale \\
   (\AA)    &     (Mm)      &     (Mm)     &    (Mm)         \\ 
\hline 
WL         &      0.0      &      -       &  -              \\
1600       &      1.2      &     0.5      &  ~$2.4 \pm0.0$  \\
171        &      4.4      &     8.2      &  $21.1 \pm 5.1$ \\
171 rms    &       -       &     4.9      &    -            \\
\hline 
\end{tabular}
\end{table}    

For the AIA data we computed the intensity as a function of height for four 15\degr\ wide sectors centered at the north, south, east and west limbs, using the images before the transit as explained above; for the south limb we avoided the region of the streamer just east of the south pole (Figure \ref{fig:limb_AIA}). 

\begin{figure}
\begin{center}
\includegraphics[width=.9\textwidth]{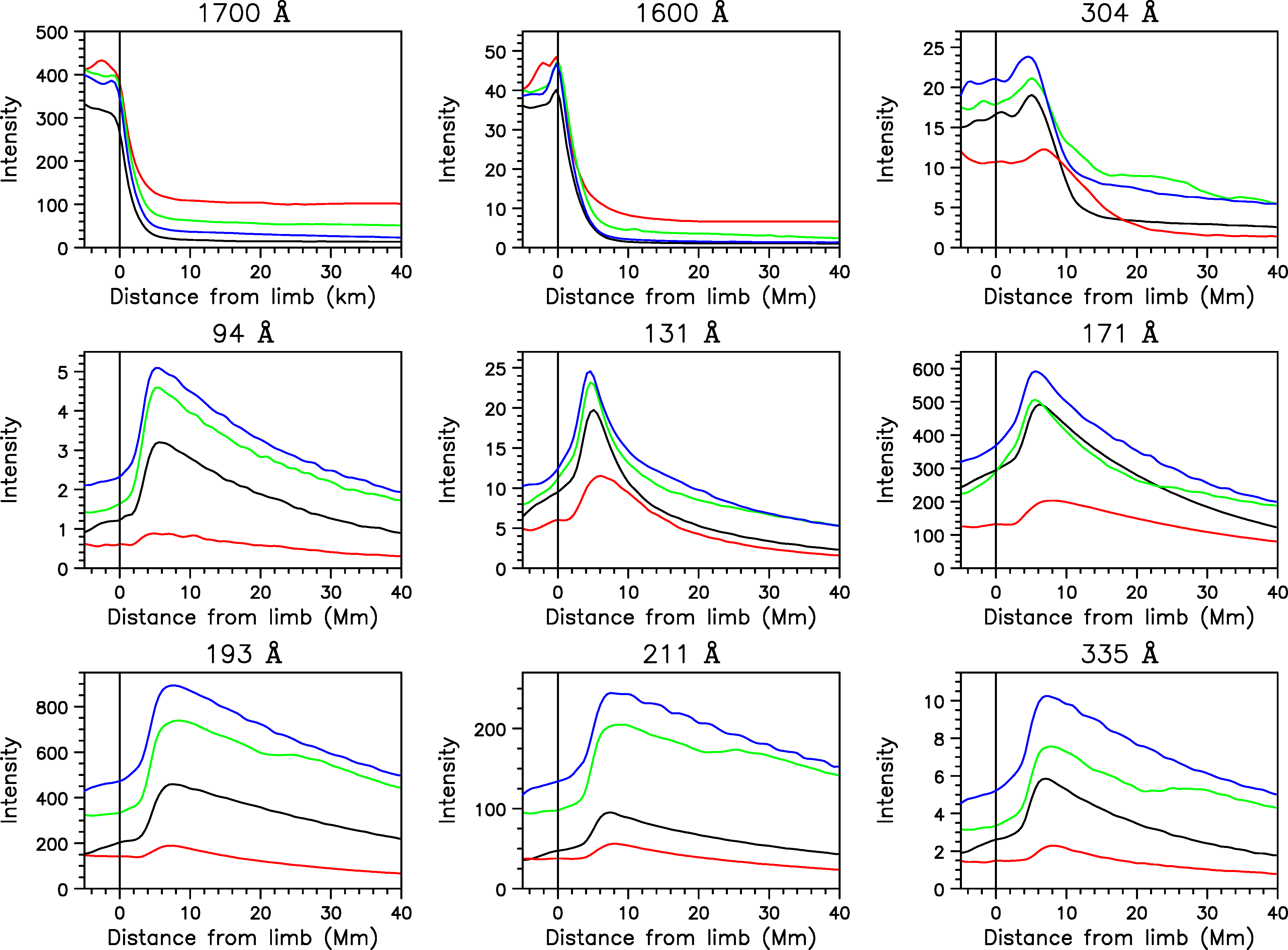}
\end{center}
\caption{The azimuthally averaged intensity, as a function of distance from the white-light limb for the nine AIA wave bands. Black is for the north limb, red for south, blue for east and green for west.}
\label{fig:cnlSDO}
\end{figure}

The intensity as a function of position near the limb is given in Figure \ref{fig:cnlSDO}, for each wavelength band. The curves show the characteristic shapes of the corresponding TRACE curves (Figure \ref{fig:cnl}), but we have important differences among the four limbs. As mentioned previously, there was some activity near the E and W limbs; as a result, we have fluctuations in the decaying part of the corresponding intensity curves, most prominent in the 211 and 335\,\AA\ bands. Beyond the S limb, there is a significant amount of scattered light in the UV bands (1600 and 1700\,\AA).

\begin{table}[h]
\caption{Limb parameters from SDO observations}
\label{Table03}
\begin{tabular}{r|c|rrrrrr|rrrrrr}
\hline 
\multicolumn{1}{c}{$\lambda$}&$\log T$&\multicolumn{6}{c}{Limb height (Mm)}&\multicolumn{6}{c}{Peak intensity height (Mm)}\\
\multicolumn{1}{c}{(\AA)}&& N & S & E& W& NS & EW &    N &    S &    E &    W & NS & EW \\
\hline
  WL &  --  &  0.0  &$-0.1$&$-0.2$& 0.0 &$-0.1$&$-0.1$&  --  &  --  &  --  &  --  &  --  &  --  \\
1700 &  --  &  0.7  &  0.7 &  0.8 & 0.7 &  0.7 & 0.8  &  --  &  --  &  --  &  --  &  --  &  --  \\
1600 &  --  &  0.9  &  0.9 &  1.5 & 1.2 &  0.9 & 1.4  &$-0.3$&$-0.4$&$-0.1$&$-0.3$&$-0.4$&$-0.2$\\
 304 & 5.00 &  8.5  & 13.7 &  7.8 & 7.0 &11.70 & 7.4  &  5.0 &  6.8 &  5.1 &  4.5 & 5.9  &  4.8 \\
  94 & 6.01 &  3.4  & 2.8  & 3.2  & 3.3 & 3.10 & 3.3  &  6.0 &  5.0 &  5.4 &  5.5 & 5.5  &  5.5 \\
 131 & 5.75 &  3.5  & 3.4  & 3.4  & 3.1 & 3.45 & 3.3  &  5.1 &  6.1 &  4.8 &  4.5 & 5.6  &  4.7 \\
 171 & 5.94 &  4.2  & 3.9  & 3.7  & 3.7 & 4.05 & 3.7  &  6.3 &  8.0 &  5.6 &  5.7 & 7.2  &  5.7 \\
 193 & 6.17 &  4.6  & 4.7  & 4.4  & 4.4 & 4.65 & 4.4  &  7.6 &  7.3 &  8.4 &  7.7 & 7.5  &  8.1 \\
 211 & 6.25 &  5.1  & 5.4  & 4.7  & 4.9 & 5.25 & 4.8  &  7.4 &  8.2 &  8.8 &  8.0 & 7.8  &  8.4 \\
 335 & 6.40 &  4.7  & 5.7  & 4.8  & 4.6 & 5.20 & 4.7  &  7.2 &  8.2 &  7.8 &  7.4 & 7.7  &  7.6 \\
\hline 
\end{tabular}
\end{table}    

The values of the characteristic parameters for all four limb regions are given in Table \ref{Table03}, where we also give the average values for the north and south, as well as for the east and west limb regions (NS and EW in the table). As in the case of TRACE, the limb height for WL, 1700, 1600 and 304\,\AA\ bands was computed from the inflection point in the decaying portion of the intensity curve (outer limb), while for the TR bands (94, 131, 171, 193, 211 and 335\,\AA) it was computed from the inflection point of the rising portion of the intensity curve (inner limb).  In the same table we give the characteristic temperature of each AIA band, from \cite{2010A&A...521A..21O} and \cite{2014SoPh..289.2377B}. For the 94\,\AA\ band we give the low temperature peak of the response curve, since we do not expect any $10^7$\,K plasma in the quiet sun; we further note that the temperature response of the 335\,\AA\ band is very extended, hence it is difficult to associate a characteristic temperature. 

As discussed in Section \ref{section:Venus}, we have an uncertainty of 0.1\arcsec\ in the position of the WL limb due to the deviation of the HMI image from a perfect circle, and this explains why the height of the WL limb in Table \ref{Table03} is not exactly zero. An additional error of 0.1\arcsec\ to 0.3\arcsec, also discussed in Section \ref{section:Venus}, comes from the accuracy of the position of Venus. We thus estimate the overall accuracy of the values in Table \ref{Table03} to be about 0.2\,Mm. On top of that, there is a dispersion of values due to the rugged form of the limb, of the order of 0.2\arcsec\ (the rms of the fit of the 1600\,\AA\ limb to a circle, see Section \ref{section:Venus}). The PSF correction had a small effect on the limb parameters of Table \ref{Table03}, the rms deviation between the two sets being 0.16\,Mm. Compared to the values obtained without the pointing corrections, the deviation was 0.26\,Mm, close to our estimate in Section \ref{section:Venus}. 

In all EUV bands, including 304\,\AA, the south limb was the weakest; it is the region near the coronal hole boundary, where spicules went higher (Section \ref{spic} and Figure \ref{fig:limb_AIA}). This asymmetry is reflected in the  304\,\AA\ limb and peak intensity heights (Table \ref{Table03}), but not in the corresponding parameters of the other bands, with the possible exception of the 94\,\AA\ band where the southern limb is 0.8\,Mm lower than the northern limb and the 335\,\AA\ band where the southern limb is 1\,Mm higher than the northern limb.

Comparing the 1600\,\AA\ and 171\,\AA\ AIA results with those from TRACE, we note that the height of the limb is very close, being only 0.2-0.3\,Mm lower in AIA. The peak intensity at 1600\,\AA\ is just inside the WL limb in TRACE and just outside in the AIA images, whereas at 171\,\AA\ the peak intensity is 1.9\,Mm higher in the TRACE images. These differences could be attributed to differences in the filter response and/or time variability.

It is a well known fact that spicules intrude well into the corona (see, {\it e.g.\/}, \citealp{1968SoPh....3..367B, 1972ARA&A..10...73B, 2012SSRv..169..181T}), thus it comes as no surprise that the height of the limb in the TR bands is well below the average spicule height. We note that the height of the 304\,\AA\ peak intensity is also above the TR band limb, by 6.4\,Mm on the average for the poles. This effect has been noted  in the past by \cite{1998ApJ...504L.127Z}, who reported a height difference of $(6.6\pm1.2)$\,Mm between the 304\,\AA\ limb and the limb in the other EIT bands at 171, 195 and 284\,\AA\ in polar regions. Still the 304\,\AA\ peak intensity is below that of most TR bands. We conclude that the plasma emitting in the 304\,\AA\ line does not absorb much of the EUV radiation ({\it c.f.} Section \ref{spic}). 

The radii of the EIT images given by \cite{1998ApJ...504L.127Z} lead to a height of 7.7\,Mm for the TR limb in polar regions, 1.3\,Mm greater than the average of out measurements  (6.4\,Mm). On the other hand, \cite{2007A&A...476..369G} measured  from EIT images heights of $(5.71\pm 0.03)$\,Mm and $(7.146\pm 0.015)$\,Mm for 304 and 171\,\AA\ respectively, with practically no solar cycle variation. The apparent contradiction with our results is due to the fact that what they actually measured was the height of peak intensity, for which the average of the values in Table \ref{Table03} is 5.4\,Mm for 304\,\AA\ and 6.4\,Mm for 171\,\AA. 

More measurements of the solar radius were compiled by \cite{2015ApJ...812...91R} for the optical and EUV and by \cite{2017SoPh..292..195M} for the microwave range. \cite{2015ApJ...798...48E} compared the position of Venus from the same transit to ephemeris data and reported radii, reduced to 1 AU, of 959.57\arcsec, 961.76\arcsec\ and 963.04\arcsec\ for HMI, 1700\AA\ and 1600\AA, respectively. These correspond to heights above the HMI limb of 1.59\,Mm for 1700\AA\ and 2.51\,Mm for 1600\AA, about a factor of two higher than our values; this  difference could be attributed to the different method used and, probably, to the fact that their measurement refers to the points of the planet's ingression and egression.
In the microwave range the situation is complicated, as mentioned in the Introduction, due to the low resolution of full-disk images and the effect of spicules. For reference we note that the radii reported by \cite{2017A&A...605A..78A} correspond to limb heights of 1.1\,Mm at 239\,GHz and 3.2\,Mm at 100\,GHz, while those of \cite{2017SoPh..292..195M} imply a limb height of 5.0\,Mm both at 200 and 400\,GHz. At longer wavelengths (17\,GHz) \cite{2004A&A...420A..1117S} gave an average polar radius of 974.4\arcsec, corresponding to a height of 10.7\,Mm.

In spite of the activity in the east and west polar regions, the prolate nature of the chromospheric limb (\citealp{1998A&A...336L..57A, 1998ApJ...504L.127Z}) is clear in the limb height values of Table \ref{Table03} at 304\,\AA, where the pole/equator height ratio is 1.6. In the TR bands it is less clear, with an average ratio, excluding the 94\,\AA\ band, of $1.08\pm0.02$ while in the peak intensity height it is only clear at 304\,\AA\ (ratio of 1.2). 

\subsection{Column Density of Neutral Hydrogen}
\begin{figure}
\begin{center}
\includegraphics[width=.9\textwidth]{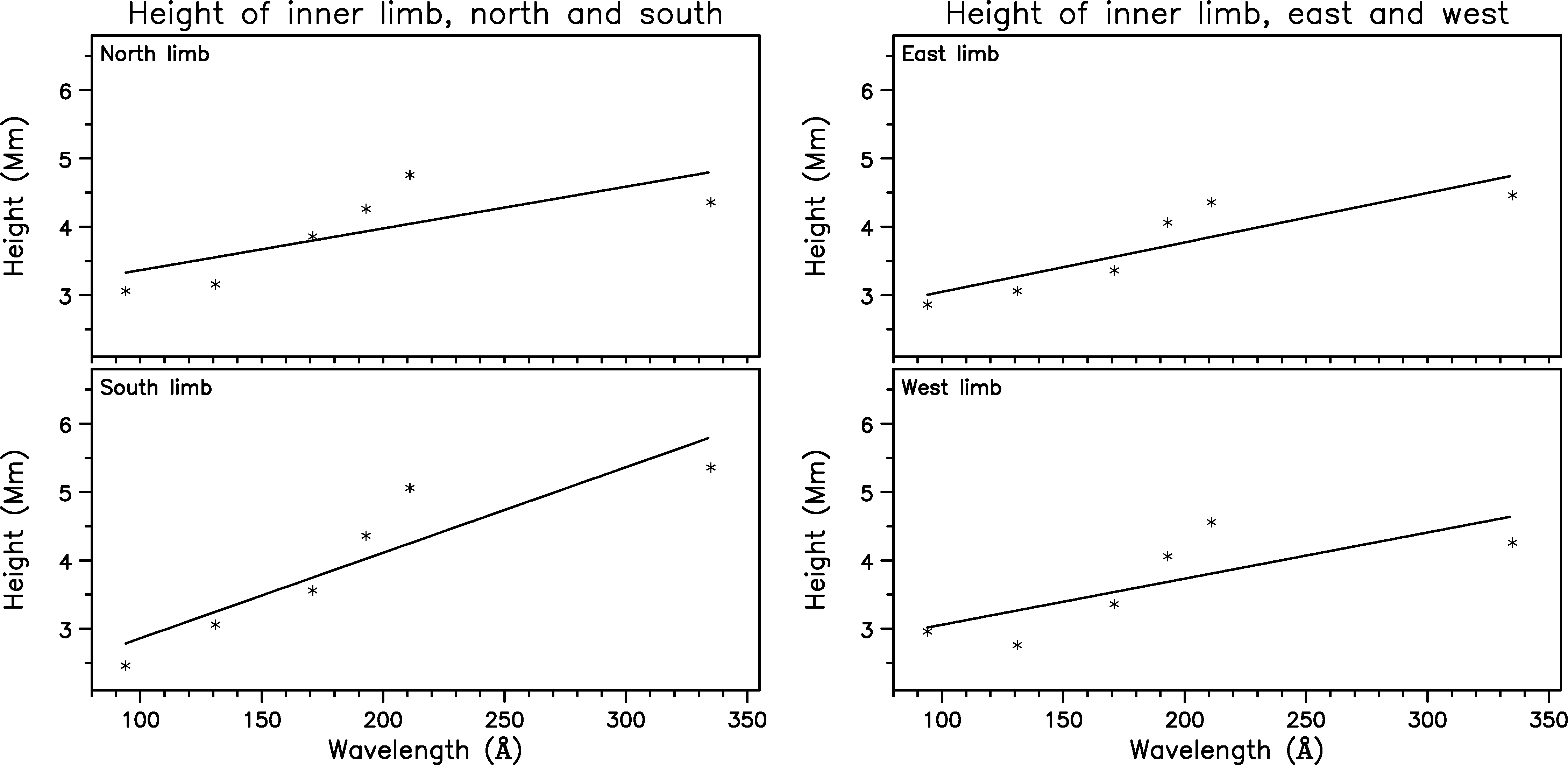}
\end{center}
\caption{The height of the inner limb with respect to the $\tau_{5000}=1$ level for the AIA TR channels as a function of wavelength. The straight lines show the result of linear regression.}
\label{fig:temp}
\end{figure}

Figure \ref{fig:temp} shows the dependence of the inner limb height on the wavelength for the TR bands. The height is from the $\tau_{\rm5000}=1$ level, which is about 340\,km below the optical limb (Table I-1 in \citealp{1976ASSL...53.....A}). The different slopes of the regression lines for the North and the South limb are due to difference in the 94\,\AA\ and 335\,\AA\ bands, as noted in Section \ref{radprof}. 

It is clear that, in all four limbs, the height increases with wavelength, which implies that the absorption also increases with wavelength. According to \cite{2013ApJ...764..165W}, see also \cite{2001SoPh..199..115C}, at wavelengths shorter than the Lyman series limit (912\,\AA), cool plasma is optically thick due to photo-ionization, principally of neutral hydrogen (H{\sc i}) and neutral or singly-ionized helium (He{\sc i} and He{\sc ii}). The absorption coefficient increases with wavelength, in conformity with our results. 

It is reasonable to assert that the optical depth of the chromospheric absorption is unity at the height of the TR inner limb, where coronal emission from behind the plane of the sky becomes visible. We can use this assertion to estimate the column density of neutral H, since
\be
\tau_\lambda=\sigma_\lambda\int_{-\infty}^\infty N_{\rm HI}(\ell) d\ell,
\ee  
where $\sigma_\lambda$ the absorption cross section, $N_{\rm H{\sc I}}(\ell)$ the density of neutral Hydrogen and the integration is carried out along the line of sight, $\ell$. Thus, for $\tau_\lambda=1$, the column density of neutral Hydrogen, $N_c$ is, simply,
\be
N_c(h_\lambda)=\int_{-\infty}^\infty N_{\rm HI}(\ell) d\ell=\sigma_\lambda^{-1}, \label{eq:colden}
\ee
where $h_\lambda$ is the height of the inner limb at the wavelength $\lambda$.

Due to the high ionization potential of He, we do not expect much of it to be ionized in the chromosphere; there should be some in the TR along our line of sight but, as the TR region is thin, we have ignored its contribution. Following \cite{2013ApJ...764..165W}, for the computation of the total cross section we used the expressions of \cite{1996ApJ...465..487V}.  For the He abundance with respect to H we adopted the value of 0.085 \citep{2007SSRv..130..105G}. From the tables of \cite{2008ApJS..175..229A} and \cite{2015ApJ...811...87A} we computed  that H is about 50\% ionized in the chromosphere ($T\sim10^4$\,K); we thus used twice that value in our computations, so that
\be
\tau_\lambda=\tau_{\lambda, \rm HI}+\tau_{\lambda, \rm HeI}
= (\sigma_{\lambda, \rm HI}+0.17\sigma_{\lambda, \rm He})N_c
\ee

\begin{figure}[h]
\begin{center}
\includegraphics[width=.6\textwidth]{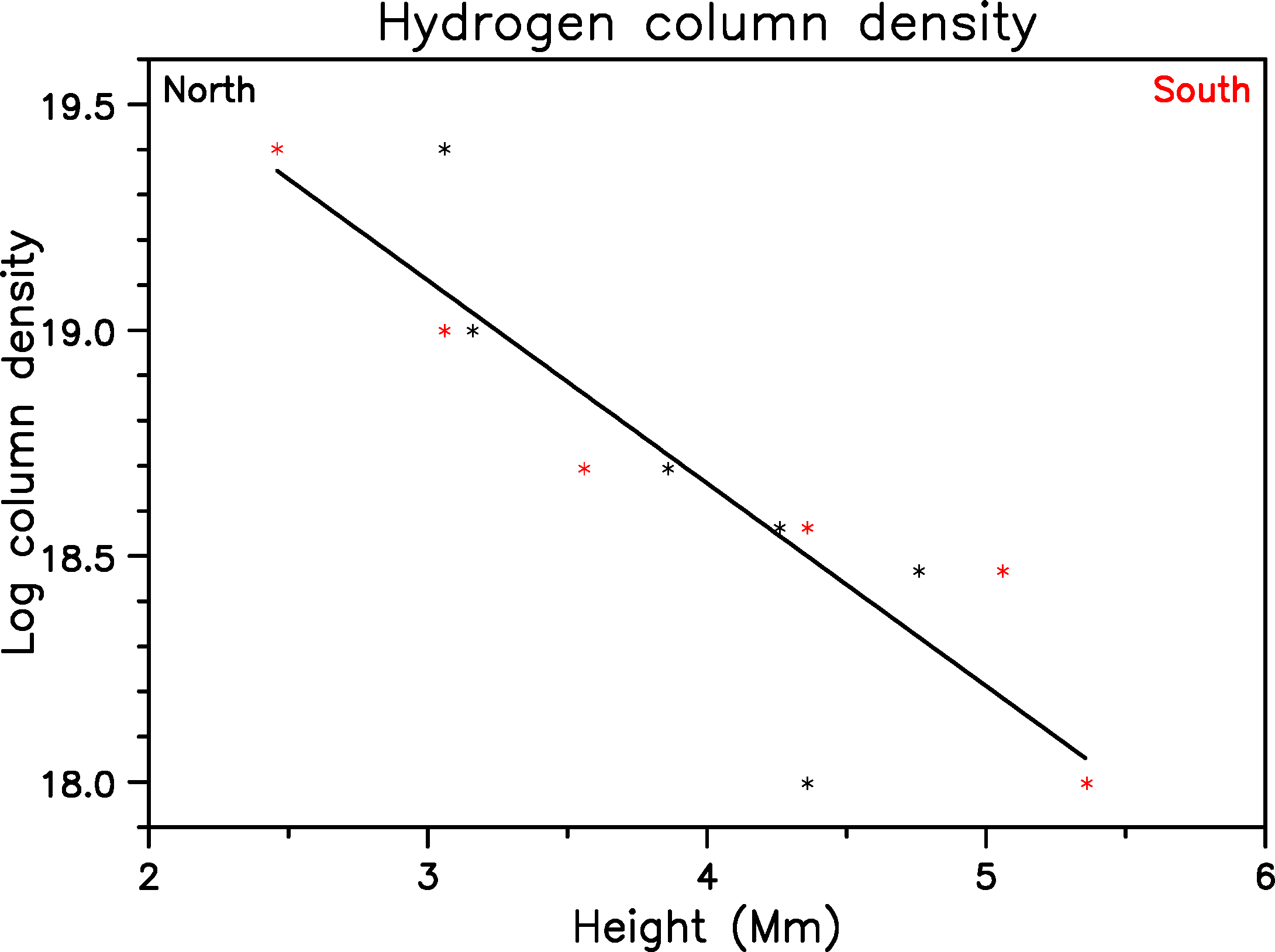}
\end{center}
\caption{Computed column density of neutral hydrogen as a function of height above the $\tau_{5000}=1$ level. Black is for the north polar region, red for the south. Both sets were used for the computation of the regression line.}
\label{fig:colden}
\end{figure}

Applying the above methodology to our measurements of  the inner limb height and combining the north and south limb regions for the five TR bands, we obtained H{\sc i} column densities from $\sim2.2\times 10^{19}$ to $\sim1.6\times10^{18}$ cm$^{-2}$, at heights of 2.5 and 5\,Mm above the $\tau_{5000}=1$ level respectively (Figure \ref{fig:colden}). In spite of the scatter, the column density shows a well-defined decrease with height, suggesting a scale of $0.97\pm0.16$\,Mm.

In order to derive the H{\sc i} density as a function of height from the H{\sc i} column density as a function of distance from the limb, let us assume that $N_{\rm HI}$ decreases exponentially with height:
\be
N_{\rm HI}(z)=N_0{\rm e}^{-z/L_{\rm HI}} \label{eq:NHI}
\ee
where $z$ is the height and $L_{\rm HI}$ is the scale height. Introducing the above expression into (\ref{eq:colden}) and after some algebra, described in detail in Appendix \ref{section:ap1}, we obtain
\be
N_c(h) \simeq N_0 \sqrt{2\pi L_{\rm HI}\,(R_{\sun}+h) } \,{\rm e}^{-h/L_{\rm HI}}
\ee
which has the same functional form as (\ref{eq:NHI}). 

We can thus compute the H{\sc i} density by dividing H{\sc i} column density by the square root factor which, since $R_{\sun} >> h$ and for the scale obtained above, has the numerical value:
\be
\sqrt{2\pi L_{\rm HI}\,(R_{\sun}+h) }=6.63\times10^9 \mbox{\,Mm}
\ee

\begin{table}[h]
\caption{Hydrogen density at 4000\,km and scale height}
\label{hyden}
\begin{tabular}{ccccl}
\hline 
  $N_c$   &  $N_{HI}$ &  $N_{H}$  & Scale & Reference\\
cm$^{-2}$ & cm$^{-3}$ & cm$^{-3}$ & Mm               \\
\hline 
$4.59\times10^{18}$&$6.91\times10^ 8$&$1.38\times10^9$&  0.97 & This work                 \\
           --      &$1.36\times10^{10}$&      --        &  1.25 & \cite{1995ApJ...453..929D}\\
           --      &       --        &$0.55\times10^9$&  4.1 & \cite{2008ApJS..175..229A}\\
           --      &       --        &$1.64\times10^9$&  5.9 & \cite{2015ApJ...811...87A}\\
\hline 
\end{tabular}
\end{table}    

Our results are summarized in Table \ref{hyden}, where we give the H{\sc i} column density and density, interpolated at a height of 4\,Mm above the $\tau_{5000}=1$ level; the total H density, $N_{H}$, is twice the H{\sc i} density, since we have assumed 50\% ionization. In the same table we give the NIXT results of \cite{1995ApJ...453..929D} from their Figure 14, the total H density from the models of \cite{2008ApJS..175..229A} and \cite{2015ApJ...811...87A} and the corresponding scale heights. The value of \cite{1995ApJ...453..929D} was computed through model fitting of the X-ray emission, using a filling factor of a few percent. An averaging over 19 angular directions was performed and the scatter of the derived density values was about an order of magnitude. The filling factor is probably the principal responsible for the large difference between their result and ours, derived for a homogeneous chromosphere. 

Our results cannot be directly compared with the H{\sc i} density of the models of \cite{2008ApJS..175..229A} and \cite{2015ApJ...811...87A} because, according to them, at 3\,Mm we are well within the TR; we thus preferred to compare the total H density. The two atmospheric models give very different values both for the density and the sale height; our density value is about twice that of the \cite{2015ApJ...811...87A} model, with a considerably smaller scale height.

\subsection{Height of the Top of the Chromosphere}
The presence of spicules does not allow a direct measurement of the average height of the top of the chromosphere. An upper limit can be deduced from the minimum height of the inner limb at 94\,\AA, which is about 3000\,km. A better estimate can be obtained from the extrapolation of the regression line of the average inner limb height versus wavelength plot (Figure \ref{fig:temp}) to $\lambda=0$, where there is no chromospheric absorption. This gives a height of $(2300 \pm 500)$\,km, a range which is consistent with the value of 2140\,km used by \cite{2008ApJS..175..229A}. 
 
We note that the limb height for the Mg{\sc ii} triplet lines at 2791.6\,\AA\ and 2798.8\,\AA, measured with respect to the 2832.02\,\AA\ continuum limb from {\it Interface Region Imaging Spectrograph} (IRIS) spectra by \cite{2018SoPh..293...20A}, are 1.7\,Mm and 2.1\,Mm respectively; these lines show no spicular structure, hence their heights place a lower limit to the height of the top of the homogeneous chromosphere.

We should bear in mind that between the top of the chromosphere and the top of 304\,\AA\ spicules we have a mixture of chromospheric, TR and coronal material, which implicates a height varying filling factor.

\subsection{Scale Height of the Coronal Emission}
We will start by computing the AIA intensity as a function of distance from the limb for an isothermal, hydrostatic corona. Let $R_j$ be the temperature response function for the AIA spectral channel $j$. The corresponding observed intensity, $I_j$, is:
\be
I_j=\int_{T_{e_1}}^{T_{e_2}}\,\varphi(T_e)\, R_j(T_e)\, d T_e  \label{Int3}
\ee
where  $T_e$ is the temperature and $\varphi(T_e)$ is the differential emission measure (DEM), defined as:
\be
\varphi(T_e)=N_e^2(\ell)\frac{d\ell}{dT_e(\ell)}, \mbox{\hspace{0.5cm} [cm$^{-5}$\,K$^{-1}$]} \label{demdef}
\ee
where $\ell$ is the distance along the line of sight and $N_e$ the electron density. 

For an isothermal corona of temperature $T_e$, the DEM at a distance $h$ from the limb is:
\be
\varphi(T_e,h)=EM(h) \delta (T_e-T_c)	\label{isophi}
\ee
where $EM$ is the emission measure (EM) is defined as:
\be
EM=\int_{\ell 1}^{\ell 2}N_e^2(T_e)d\ell=\int_{T_{e1}}^{T_{e2}}\varphi(T_e)dT_e, \mbox{\hspace{0.5cm} [cm$^{-5}$]} \label{EMdef}
\ee
Substituting in (\ref{Int3}) we get
\be
I_j(h)=EM(h) R_j(T_c)
\ee
From which it is obvious that the form of $I_j$ should be the same for all AIA channels, the values differing  by the constant factor $R_j(T_c)$.

Adopting a hydrostatic and isothermal coronal model with constant gravity,
\be
N_e(h)=N_o{\rm e}^{-h/L_{\rm cor}} \label{eq:NH}
\ee
where $L_{\rm cor}$ is the coronal scale height, (\ref{EMdef}) gives:
\be
EM(h)
=N^2_o\int_{-\infty} ^{\infty}{\rm e}^{-2h/L_{\rm cor}} d \ell
\label{IntEM2}
\ee
Equation (\ref{eq:NH}) is of the same form as (\ref{eq:NHI}), with $z/L_{HI}$ replaced by $2h/L_{\rm cor}$, and (\ref{IntEM2}) evaluates to
\be
EM(h) \simeq N_0 \sqrt{\pi L_{\rm cor}\,(R_{\sun}+h) } \,{\rm e}^{-2h/L_{\rm cor}}
\ee
The important thing to notice is that the scale height of the AIA emission is half of the hydrostatic scale height (see also \citealp{2001ApJ...550..475A}).

\begin{table}[h]
\caption{Intensity scale height from AIA images}
\label{intscale}
\begin{tabular}{r|ccc}
\hline 
$\lambda$&\multicolumn{3}{c}{$L_{\rm cor}$}~(Mm) \\
         & North & South & Average \\
\hline 
131      & 26.5  & 24.7  &  25.6   \\
171      & 24.6  & 32.2  &  28.4   \\
 94      & 28.2  & 32.7  &  30.5   \\
335      & 31.2  & 39.6  &  35.4   \\
193      & 40.2  & 34.8  &  37.5   \\
211      & 46.6  & 41.1  &  43.9   \\
\hline 
\end{tabular}
\end{table}    

Contrary to the above prediction, the intensity scale height, measured by fitting the observed intensity curves in the height range of 20 to 40\,Mm, is not the same for all AIA TR channels (Table \ref{intscale}). As a matter of fact, the scale height in the 193 and 211\,\AA\ bands is almost twice that of the 131 and 171\,\AA\ bands, a difference that can also be seen in Figure \ref{fig:cnl}. This could be the result of insufficient scattered light correction and/or a non-isothermal corona (see, {\it e.g.} \citealp{1998A&A...336L..90D, 2001ApJ...550..475A}); the increased scale height for AIA bands sensitive to high temperatures is suggestive of the latter. There is no significant north-south asymmetry, and the average intensity scale height is $(33.5\pm 6.8)$\,Mm, which translates to the reasonable value for the coronal temperature at the poles of $(1.24\pm0.25)\times10^6$\,K. We add that, at 171\,\AA, we measured with TRACE a scale height smaller by $\sim7$\,Mm (Table \ref{Table02}). 

A more precise analysis should attempt to model the AIA emission, taking into account the chromospheric absorption, the height varying filling  factor, the temperature/density structure of the atmosphere and the AIA response. Another possible approach would be to compute the DEM from the AIA intensities and invert it using the method of \cite{2019SoPh..294...23A} to obtain the temperature/density structure; this method could be used for quiet regions on the disk, as well as beyond the limb. We intend to investigate these possibilities in future works.

\section{Summary and Conclusions}\label{section:disc}
Using images from the Mercury and Venus transits, we computed significant pointing corrections for TRACE images and very small ones for SDO. This allowed us to align properly the images in the various spectral bands, with an accuracy of about 0.2\,Mm, and study the spicular structures (in emission and in absorption) beyond the limb; we also computed the limb height, the height of maximum emission, as well as the intensity scale height. AIA images were corrected for scattered light using the PSF of \cite{2013ApJ...765..144P}.

At low heights the chromosphere is opaque to EUV radiation. Coronal emission from behind the plane plane of the sky appears near the position of the inner limb, at a height which increases with wavelength; this implies that the chromospheric absorption also increases with wavelength, as expected for bound-free absorption due to neutral hydrogen and neutral or singly-ionized helium. At that height, absorption features in the high temperature bands are similar, but not identical, to spicules in the 1600\,\AA\ band and the Ca{\sc ii} H-line band of SOT. 

Spicules at 304\,\AA\ extend much higher than in the 1600\,\AA\ band, the intensity variation along the limb of the latter at a height of 7.5\arcsec\ being similar to the 304 intensity at 14.5\arcsec. Tall spicules in 304\,\AA\ do not appear in absorption in the TR bands, moreover the height of 304\,\AA\ peak emission is above the inner limb of the TR channels; we conclude that the 304 emitting plasma is optically thin to the EUV radiation. We also found that spicules were significantly higher above the south polar region, located near the boundary of a coronal hole, than above the north polar region. Our measurements confirmed the prolate nature of the 304\,\AA\ limb; the effect is less pronounced or absent in the high temperature AIA bands.

Assuming that the optical depth of the chromospheric absorption in the TR bands is unity at the height of the inner limb, and attributing the absorption to neutral hydrogen and helium, we computed the neutral hydrogen column density as a function of height; we obtained values between $\sim2.2\times 10^{19}$ and $\sim1.6\times10^{18}$\,cm$^{-2}$, at heights between 2.5 and 5\,Mm above the $\tau_{5000}=1$ level. From these results and for a homogeneous chromosphere, we computed a  hydrogen density of $6.9\times10^8$\,cm$^{-3}$ at a height of 4\,Mm and a scale height of 0.97\,Mm.

We computed the scale height of the coronal emission for a homogeneous and isothermal corona and found that it is one half of the density scale height; moreover, we found that it should have the same value for all wavelength bands that are sensitive to coronal emission.  However, the observed scale height varies considerably among these wavelength bands, most probably due to the failure of the isothermal hypothesis. The average scale height leads to a temperature of $(1.24\pm0.25)\times10^6$\,K for the polar corona.

AIA images provide important diagnostics of the quiet TR and low corona. As the pointing corrections are small, the present analysis could be extended to any set of SDO images and thus check for variations with the solar cycle. In addition, more detailed analyses could use the differential emission measure derived from AIA images and thus extend the analysis in quiet regions on the solar disk. Moreover, under favorable conditions, the absorption of EUV radiation by individual spicules could be measured, and this would provide important information for modeling the physical conditions in spicules.

\begin{acks}
The authors gratefully acknowledge use of data from the TRACE, Hinode and SDO (AIA and HMI) databases. We also want to thank S. Patsourakos and A. Nindos for comments on the manuscript and suggestions.
\end{acks}

\medskip\noindent{\footnotesize {\bf Disclosure of Potential Conflicts of Interest} The authors declare that they have no conflicts of interest.}

\appendix
\section{Relation Between the Column Density, $N_c(h)$, and the Density, $N(z)$}\label{section:ap1}
Doing a simple geometric transformation we can write (\ref{eq:NHI}) in the form:
\be
N_{\rm HI}(\ell)=N_0 \exp \left(-\frac {\sqrt{\ell^2+r^2}-R_{\sun}} {L_{\rm HI}}\right)
\ee
where $\ell$, as before, is along the line of sight, $r$ is the distance from the center of the sun and $R_{\sun}$ the solar radius. The column density is then, from (\ref{eq:colden})
\be
N_c(r)=N_0{\rm e}^{\frac{R_{\sun}}{L_{\rm HI}}}\int_{-\infty}^\infty{\rm e}^{\left(-\frac {\sqrt{\ell^2+r^2}} {L_{\rm HI }}\right)}d\ell
\ee
The integral can be computed using  the expression 3.461/6 in the table of \cite{2007tisp.book.....G}, p.364:
\be
\int_0^{\infty} \exp (-a\sqrt{x^2+b^2}\,) dx=b K_1 (ab)
\ee
where $K_1$ is the modified Bessel function of the first order. For $a=1/L_{\rm H {\sc i}}, b=r$, this gives
\be
N_c(h)=2N_0{\rm e}^{\frac{R_{\sun}}{L_{\rm HI}}}\, r K_1\left(\frac{r}{L_{\rm HI}}\right) \label{colden2}
\ee

Since $r=R_{\sun}+h$, where h is the distance from the limb, the argument of $K_1$ in (\ref{colden2}) is large; we can thus retain the first order term in the expansion of $K_\nu(z)$ for large $z$ (expression 8.451/6 from p. 920 of \citealp{2007tisp.book.....G}),
\be
K_\nu(z)\simeq\sqrt{\frac{\pi}{2z}}{\rm e}^{-z},
\ee
and, substituting in (\ref{colden2}), we obtain, finally:
\be
N_c(h) \simeq 
N_0 \sqrt{2\pi L_{\rm HI}\,(R_{\sun}+h) } \,{\rm e}^{-h/L_{\rm HI}}
\ee

\end{article} 
\end{document}